\documentclass[12pt]{article} 
\usepackage{latexsym,amssymb,epsfig}  
\newcommand{\bm}[1]{\mbox{\boldmath$#1$}} 
\def\0{\phantom{0}} 
 
\begin{document} 
\pagenumbering{arabic} 
\baselineskip25pt 
 
\begin{center} 
{\bf \large Thermodynamic Models for Vapor-Liquid Equilibria of Nitrogen+Oxygen+Carbon Dioxide at Low Temperatures} \\ 
\end{center}


\renewcommand{\thefootnote}{\fnsymbol{footnote}}

\noindent
Jadran Vrabec$^1$\footnote{corresponding author, Tel.: +49-711/685-66107, Fax:
+49-711/685-66140, Email: vrabec@itt.uni-stuttgart.de}, Gaurav Kumar Kedia$^1$, Ulrich Buchhauser$^2$, Roland Meyer-Pittroff$^3$, Hans Hasse$^1$ \\
\renewcommand{\thefootnote}{\arabic{footnote}}
{\rm{$^1$ Institut f\"ur Technische Thermodynamik und Thermische Verfahrenstechnik, Universit\"at Stuttgart, D-70550 Stuttgart, Germany}} \\
{\rm{$^2$ Lehrstuhl f\"ur Rohstoff- und Energietechnologie, Technische Universit\"at M\"unchen, D-85350 Freising-Weihenstephan, Germany}} \\
{\rm{$^3$ Competence Pool Weihenstephan, Technische Universit\"at M\"unchen, D-85354 Freising-Weihenstephan, Germany}}

\begin{abstract} 
\baselineskip21pt
For the design and optimization of CO$_2$ recovery from alcoholic fermentation processes by distillation, models for vapor-liquid equilibria (VLE) are needed. 
Two such thermodynamic models, the Peng-Robinson equation of state (EOS) and a model based on Henry's law constants, are proposed for the ternary mixture N$_2$+O$_2$+CO$_2$. 
Pure substance parameters of the Peng-Robinson EOS are taken from the literature, whereas the binary parameters of the Van der Waals one-fluid mixing rule are adjusted to experimental binary VLE data. 
The Peng-Robinson EOS describes both binary and ternary experimental data well, except at high pressures approaching the critical region. 
A molecular model is validated by simulation using binary and ternary experimental VLE data. 
On the basis of this model, the Henry's law constants of N$_2$ and O$_2$ in CO$_2$ are predicted by molecular simulation. 
An easy-to-use thermodynamic model, based on those Henry's law constants, is developed to reliably describe the VLE in the CO$_2$-rich region. 
\end{abstract}

\clearpage 
 
\section{Introduction}  
The recovery of CO$_2$, produced during alcoholic fermentation, is an economically and ecologically interesting process for breweries. 
CO$_2$, which is the main component of the fermentation flue gas, can be collected and purified to be used as an auxiliary material during the production process or for other purposes. 
E.g., CO$_2$ is used to avoid contact of the beer product with O$_2$ from the atmosphere to minimize oxidation processes and flavor derogation.
It is needed in this case with a high purity, particularly the O$_2$ concentration has to be below 5 ppm
\cite{neue1,neue2}.

CO$_2$ originates from the anaerobic metabolism of yeast during fermentation processes. 
Yeast, mostly saccharomyces cerevisiae uvarum varians carlsbergensis, ferments the hexoses glucose and fructose as well as the disaccharides saccharose and maltose and the trisaccharides maltotriose. 
The fermentation produces per liter of beer about 42 g CO$_2$. Wort solves about 4 g/l, thus 38 g/l are released and may predominantly be recovered.
Fermentation by-products and components from the atmosphere in the fermentation tank contaminate the emerging CO$_2$ and can be separated by different cleaning steps via activated carbons and silica gels
\cite{neue4}. 
The permanent gases N$_2$ and O$_2$ from the atmosphere, however, are the main obstacles in the recovery of CO$_2$. 
Available experimental data indicates low solubilities of O$_2$ and N$_2$ in
liquid CO$_2$ at temperatures below $-35~^{\circ}$C, which allow a sufficiently high purification \cite{neue5}.

N$_2$ and O$_2$ can be separated from CO$_2$ by distillation, recovering up to $66$ \% (i.e. $25$ g/l) of the produced CO$_2$.
Common one-stage cooling devices used in breweries, mostly with NH$_3$ as working agent, are unable to provide the required low temperatures
down to $-55~^{\circ}$C \cite{neue6}.
Therefore, an additional low temperature stage has to be considered. 
CO$_2$ itself is an appropriate cooling agent for such devices due to its volumetric refrigerating capacity and the optimal pressure range at the required temperatures.
In a pilot plant at the Flensburger Brauerei, Emil Petersen GmbH \& Co. KG (Germany), which is discussed in \cite{MBAA}, a cascade cooling device was added. 
After liquefaction and super-cooling, the recovered CO$_2$-rich liquid is stored into an interim tank. 
Subsequently, the liquid mixture is heated by three heating sections to remove the super-cooling and to boil out the permanent gases N$_2$ and O$_2$. 
The gaseous fraction is dissipated to avoid a re-solution.

For the design and optimization of such recovery and cleaning processes, vapor-liquid equilibrium (VLE) data for the ternary mixture O$_2$+N$_2$+CO$_2$ at temperatures between $-55$ and $-20~^{\circ}$C are needed. 
The focus of the present work is the assessment of the available experimental VLE data and the development of two thermodynamic models for that purpose. Particular attention is given to the CO$_2$-rich region.
Beside the classical approach with the Peng-Robinson equation of state (EOS), molecular modelling and simulation were used to develop an easy-to-use model based on Henry's law constants. 
It should be noted that the molecular approach is very much suitable to predict mixing properties for a wide variety of fluids.

\section{Experimental Data} 
Experimental VLE data of the ternary mixture 
N$_2$+O$_2$+CO$_2$ are available only from two publications, Zenner et al.  
\cite{zenn} and Muirbrook et al. \cite{muir_oc_noc}. 
The later \cite{muir_oc_noc} deals exclusively with the 0$~^{\circ}$C isotherm, thus Zenner et al. \cite{zenn} is the only ternary source within the regarded temperature range. 
In \cite{zenn}, the VLE has been measured at the temperature $-$40.3$~^{\circ}$C and pressures of 5.17, 6.90 and 12.69 MPa with CO$_2$ liquid mole fractions ranging from 0.73 to 0.92.
At $-$55$~^{\circ}$C pressures of 6.90, 10.35, and 13.10 MPa have been investigated with CO$_2$ liquid mole fractions ranging from 0.69 to 0.87. 
These are high pressure VLE so that the very CO$_2$-rich region is not covered.  
 
Regarding the three binary subsystems, the available experimental data base is much better and extensive VLE data can be found. 
For N$_2$+O$_2$, which consists of the two lower boiling components of the present ternary mixture, 15 publications are available [8-22].
Both components have very low critical temperatures, i.e. $-147.05$~$^{\circ}$C (N$_2$) and  $-$118.57$~^{\circ}$C (O$_2$), so that the binary VLE lies considerably lower than the temperature range of interest as well. 
 
The VLE of the binary subsystem N$_2$+CO$_2$ has been investigated in 21 publications [6,7,23-41].
CO$_2$ has a much higher critical temperature of 30.98$~^{\circ}$C and experimental binary VLE data can be found in the regarded temperature range in 12 publications [25-36].
 
Also for the third subsystem, i.e. O$_2$+CO$_2$, sufficient VLE data is available [6,31,42-44].
In this case, numerous data points are known in the relevant temperature range from $-$55 to $-$20$~^{\circ}$C \cite{zenn, fred, fred66_oc} as well. 
 
In the following, a subset of the experimental data was used to adjust and validate the thermodynamic models developed in this work, i.e. \cite{dodg_no, kric_no , dodg1_no} for N$_2$+O$_2$, \cite{ zenn} for N$_2$+CO$_2$, \cite{fred} for O$_2$+CO$_2$, and \cite{zenn} for the ternary mixture N$_2$+O$_2$+CO$_2$. 
For N$_2$+O$_2$ and N$_2$+CO$_2$ those data sets were selected, which contain a larger number of data points.

\section{Molecular Model} 
 
To describe the intermolecular interactions in the ternary mixture, effective state independent pair potentials were used here, which implies that many-body interactions were neglected.
For this purpose, the two-centre Lennard-Jones plus point quadrupole 
(2CLJQ) pair potential was employed \cite{vraquad}.   
It is composed of two identical Lennard-Jones sites a distance $L$ apart (2CLJ) and an axial point-quadrupole of momentum $Q$ placed in the geometric centre of the molecule.
The intermolecular potential writes as 
\begin{eqnarray} 
u_{\rm 2CLJQ}(\bm{r}_{ij},\bm{\omega}_i,\bm{\omega}_j) = \sum_{a=1}^{2} \sum_{b=1}^{2} 4\epsilon \left[ \left( \frac{\sigma}{r_{ab}} \right)^{12} - \left( \frac{\sigma}{r_{ab}} \right)^6 \right] 
+\frac{3}{4}\frac{Q_iQ_j}{\left|\bm{r}_{ij}\right|^5} f_{\rm Q}\left(\bm{\omega}_i,\bm{\omega}_j\right). 
\label{uQ} 
\end{eqnarray} 
Herein, ${\bm r}_{ij}$ is the centre-centre distance vector of two molecules $i$ and $j$, $r_{ab}$ is one of the four Lennard-Jones site-site distances; $a$ counts the two sites of molecule $i$, $b$ counts those of molecule $j$. 
The vectors ${\bm \omega}_i$ and ${\bm \omega}_j$ represent the orientations of the two molecules $i$ and $j$. 
$f_{\rm Q}$ is a trigonometrical function depending on these molecular orientations, cf. Gray and Gubbins \cite{gray84}. 
The Lennard-Jones parameters $\sigma$ and $\epsilon$ represent size and energy, respectively. 
In total, the 2CLJQ model has four model parameters: $\sigma$, $\epsilon$, $L$, and $Q$. 
These parameters have been adjusted to VLE for numerous pure fluids in prior work \cite{vraquad}.  
Table \ref{2CLJDmodels} summarizes the parameters of the three pure fluid molecular models considered here. 
 
On the basis of existing models for pure fluids, molecular modelling of mixtures reduces to specifying the interaction between unlike molecules. 
Following prior work \cite{vlemix}, a modified Lorentz-Berthelot combining rule with one adjustable binary interaction parameter $\xi$ was used for each unlike Lennard-Jones interaction 
\begin{eqnarray} 
\sigma_{\rm AB}= {\frac{\sigma_{\rm A} + \sigma_{\rm B}}{2}}, \label{sig} 
\end{eqnarray} 
\begin{eqnarray} 
\epsilon_{\rm AB}= \xi \cdotp {\sqrt{\epsilon_{\rm A} \cdotp \epsilon_{\rm B}}}. \label{episi} 
\end{eqnarray} 
Table \ref{xi} summarizes the three binary interaction parameters needed for the ternary mixture model that were taken from \cite{vlemix}. 
The interaction between the quadrupolar sites is treated in a physically straightforward way without the use of binary parameters. 
It was shown in \cite{vlemix} that this ternary model yields an accurate description of the thermodynamic properties of this mixture. 
It can readily be used to predict a wide range of thermodynamic properties such as Henry's law constants as discussed below.
 
To calculate the VLE on the basis of this molecular model by simulation, the Grand Equilibrium method was applied here. 
The description of this method can be found elsewhere \cite{ge} and is not repeated.
Only the technical details of the present calculations are concisely given in the following. Molecular dynamics simulations for the liquid phase were performed in the isobaric isothermal ($N\!pT$) ensemble, using Anderson's barostat \cite{andersen1} and isokinetic velocity scaling \cite{allen1}. 
A total of 864 molecules were placed initially in a fcc lattice configuration in a cubic simulation volume. 
Depending upon the density of the state point, the reduced membrane mass parameter for the barostat $Mm$/$\sigma^4$ was chosen from $10^{-3}$ to 
$10^{-6}$, where $m$ is the molecular mass. 
The intermolecular interactions were evaluated explicitly up to a cut-off radius of $5\sigma$ and standard long range corrections were used, employing angle averaging as proposed by Lustig \cite{lustig}. 
 
For the vapor phase, pseudo-grand canonical (p-${\mu}VT$) Monte-Carlo simulations were performed. The cut-off radius was also $r_c=5\sigma$ and the long range corrections were considered. 
The maximum displacement was set to 5\% of the simulation box length, which was chosen to yield on average 300 to 400 molecules in the volume. 
After 1 000 initial cycles in the canonical ($NVT$) ensemble starting from a fcc lattice, 9 000 equilibration cycles in the p-${\mu}VT$ ensemble were performed.
One cycle is defined here to be a number of attempts to displace and rotate molecules equal to two times the actual number of molecules plus three insertion and three deletion attempts. 
The length of the production run was 100 000 cycles.
In this way, VLE data on the basis of the molecular model were generated. 
The results from simulation are presented in section \ref{rnd} and validated by comparison to experimental data.

\section{Peng-Robinson Equation of State} 
 
Cubic EOS offer a compromise between generality and simplicity that is suitable for numerous purposes. 
They are excellent tools to correlate experimental data and are therefore often used for many technical applications. 
In the present work, the Peng-Robinson EOS with the Van der Waals one-fluid mixing rule was adjusted to binary experimental data and validated regarding the ternary mixture. 
The Peng-Robinson EOS \cite{vann} is defined by   
\begin{eqnarray} 
p= {\frac{RT}{v-b}} - {\frac {a}{v(v+b)+b(v-b)}},\label{pr} 
\end{eqnarray} 
where the temperature dependent parameter $a$ is defined by 
\begin{eqnarray} 
 a=\left(0.45724~ \frac{R^2{T_c}^2}{p_c}\right)  
\left[1+\left(0.37464+1.54226~\omega-0.26992~\omega^2 \right)\left(1-\sqrt{\frac{T}{T_c}}\right)\right]^2.  
\label{alpha}\label{a} 
\end{eqnarray} 
The constant parameter $b$ is  
\begin{eqnarray} 
 b=0.07780~{\frac {R{T_c}}{p_c}}.\label{b} 
\end{eqnarray} 
Therein, critical temperature $T_c$, critical pressure $p_c$, acentric factor $\omega$, and the ideal gas constant $R$ of the pure substances are needed, cf. Table \ref{PRpure}. 
The values were taken from \cite{refprop}. 
 
To apply the Peng-Robinson EOS to mixtures, mixed parameters $a_m$ and $b_m$ have to be defined. For this purpose a variety of mixing rules can be found in the literature. 
Here, the Van der Waals one-fluid mixing rule \cite{vann} was chosen. 
It defines the temperature dependent parameter of the mixture as
\begin{eqnarray} 
a_m=\sum_{i} \sum_{j} x_ix_ja_{ij}.\label{am} 
\end{eqnarray} 
The indices $i$ and $j$ denote the components, with 
\begin{eqnarray} 
a_{ij}={\sqrt{a_ia_j}}(1-k_{ij}),\label{aij} 
\end{eqnarray} 
where $k_{ij}$ is an adjustable binary parameter. The constant parameter of the mixture is defined as 
\begin{eqnarray} 
b_m=\sum_{i}  x_i b_i.\label{bm} 
\end{eqnarray} 
This classical thermodynamic model was used to fit the experimental VLE data of the three binary subsystems, i.e. N$_2$+O$_2$, O$_2$+CO$_2$, and N$_2$+CO$_2$. 
The three binary parameters $k_{ij}$ were adjusted to the same experimental data as the binary parameters of the molecular model $\xi$. 
It turned out that temperature independent $k_{ij}$ values are sufficient in all cases in the regarded range of states, cf. Table \ref{binpar}. 
The results of the Peng-Robinson EOS are presented in section \ref{max} and validated by comparison to experimental data. 
 
\section{Henry Model} 
\label{henry} 
Molecular simulation allows the prediction of Henry's law constants on the basis of a given molecular model straightforwardly. 
Different approaches have been proposed in the literature, e.g. \cite{rich, mura}. 
The Henry's law constant is related to the residual chemical potential of the solute $i$ at infinite dilution ${\mu_i}^{\infty}$ 
\cite{shin} 
\begin{eqnarray} 
H_i={\rho}k_{\rm B}T \exp{({\mu_i}^{\infty}/(k_{\rm B}T))},\label{hsim} 
\end{eqnarray} 
where $k_{\rm B}$ is the Boltzmann constant, $T$ the temperature, and $\rho$ the density of the solvent.

\begin{figure}[ht] 
\epsfig{file=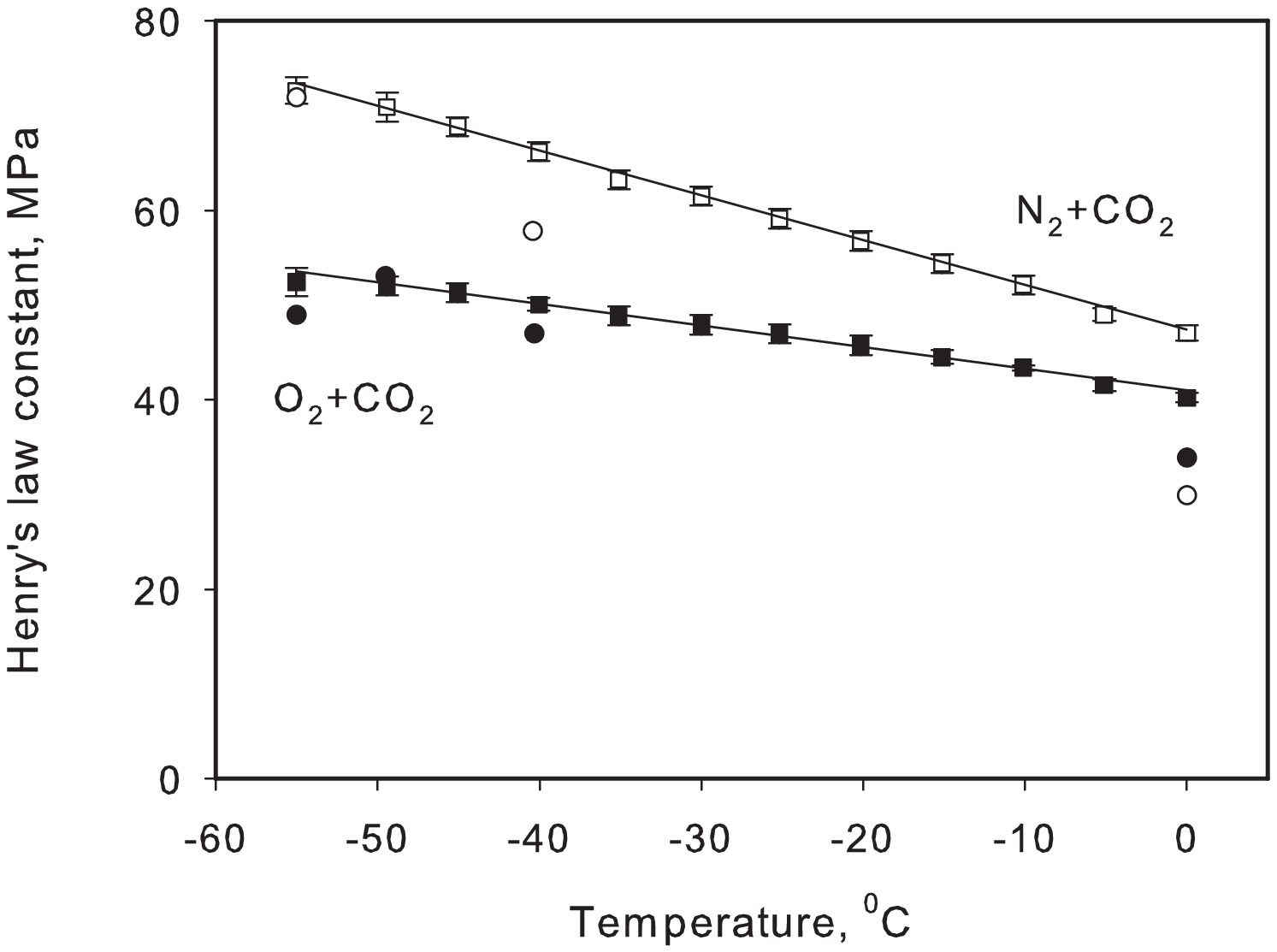,scale=0.95} 
\bigskip 
\bigskip 
\bigskip 
\bigskip 
\caption[Henry's law constants over temperature: {\large $\bullet$,\large 
$\circ$} Experimental data \cite{zenn, yori, fred}, {$\blacksquare$, $\square$} Simulation results,      
--- fit to simulation results, cf. equations 
(\ref{hn2}) and (\ref{ho2}).]{Vrabec et al.} 
\label{f1} 
\end{figure}

For the calculation of Henry's law constants, a series of simulations, ranging from 0 to $-$55$~^{\circ}$C, were performed with at intervals of 5$~^{\circ}$C.
To evaluate ${\mu_i}^{\infty}$, Widom's test insertion method \cite{wido} was used. 
Therefore, 3456 test molecules representing the solute $i$ were inserted after each time step at random positions into the liquid solvent and the potential energy between the solute test molecule and all solvent molecules ${\psi}_i$ was evaluated within the cut-off radius  
\begin{eqnarray} 
{\mu_i}^{\infty}=k_{\rm B}T{\langle}V\exp({-\psi_i/(k_{\rm 
B}T)){\rangle}}/{\langle}V{\rangle},\label{mu} 
\end{eqnarray} 
where the brackets represent the ensemble average in the $NpT$ ensemble.  
Note that appropriate long range corrections \cite{allen1} have to be applied.  
The residual chemical potential at infinite dilution from this procedure is solely attributed to the unlike solvent-solute interactions. 
The mole fraction of the solute in the solvent is exactly zero, as required for infinite dilution, since the test molecules are instantly removed after the potential energy calculation. 
Simulations were performed at a specified temperature and the according pure substance vapor pressure of CO$_2$. 
The results for the Henry's law constants $H_i$ are given as functions of the temperature as shown in Figure \ref{f1}. 
Linear functions were found to be sufficient to fit the data.  
The resulting equations are for N$_2$ in CO$_2$    
\begin{eqnarray} 
H_{\rm N2}/{\rm MPa}=178.21-0.47998~T/{\rm K}, \label{hn2} 
\end{eqnarray} 
and for O$_2$ in CO$_2$ 
\begin{eqnarray} 
H_{\rm O2}/{\rm MPa}=104.26-0.23214~T/{\rm K}. \label{ho2} 
\end{eqnarray} 
It can be seen in Figure \ref{f1} that the predicted Henry's law constant of O$_2$ in CO$_2$ agrees well with the experimental data \cite{zenn, yori, fred}, especially in the low temperature region. 
But a considerable scatter of experimental data has to be noted. 
For N$_2$ in CO$_2$ systematic deviations between simulation and experiment were found. 
At $-$55$~^{\circ}$C the agreement is very good, but with increasing temperature the data sets diverge, where simulation yields higher values. 
The deviation is 14\% at $-$40.3$~^{\circ}$C. 
 
The classical approach to model VLE on the basis of Henry's law constants includes the activity coefficient $\gamma_i$ and the fugacity coefficient $\phi_i$. 
The phase equilibrium condition for the two low boiling components $i$ = N$_2$, O$_2$ is then given by \cite{vann} 
\begin{eqnarray} 
H_i~\exp{\left\{ \frac{1}{RT}\int_{p^s_{\rm CO2}}^p v_i^{\infty} dp\right\} }~x_i~{\gamma}_i=p~y_i~{\phi}_i.\label{hi} 
\end{eqnarray} 
Here, $x_i$ and $y_i$ are the mole fractions in the saturated liquid and vapor,
respectively, and $v_i^{\infty}$ is the partial molar volume of the solute at infinite dilution. 
The exponential term, known as Krichevski-Kasarnoski correction, accounts for the higher pressure of the liquid mixture compared to the pure solvent vapor pressure. 
In the pressure range of interest, its influence is small so that this correction term was set to unity. 
 
For the solvent CO$_2$, the equilibrium condition includes the pure substance vapor pressure ${p^s}_{\rm CO2}$ instead of Henry's law constant 
\begin{equation} 
{p^s}_{\rm CO2}~x_{\rm CO2}~{\gamma}_{\rm CO2}=p~y_{\rm CO2}~{\phi}_{\rm CO2}.\label{ps} 
\end{equation} 
Therefore, a correlation for ${p^s}_{\rm CO2}$ \cite{wagnerps} was taken from the literature 
\begin{equation} 
\ln({p^s}_{\rm CO2}/p_{c})={{\sum_{i=1}^{4} A_i(1-(T/T_c))^{n_i}}/(T/T_c)},\label{corco2} 
\end{equation} 
where the parameters $A_i$ and $n_i$ are listed in Table \ref{an}.

With the previously adjusted Peng-Robinson EOS, activity coefficients were calculated in the temperature and composition range of interest. 
It was found that they are between 1 and 1.001 and were thus set to unity in the present Henry model. 

\begin{figure}[ht] 
\epsfig{file=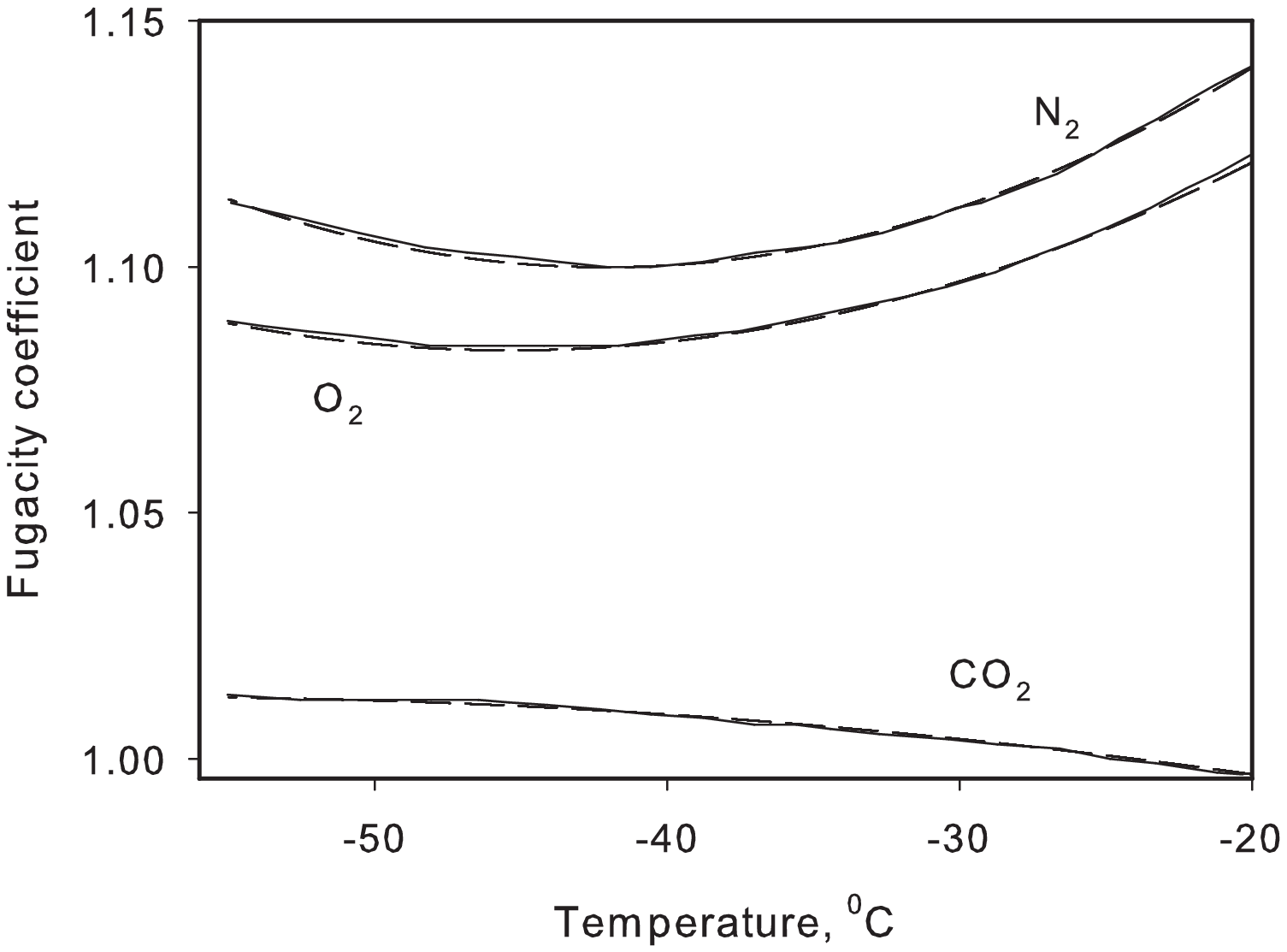,scale=0.95} 
\bigskip 
\bigskip 
\bigskip 
\bigskip 
\caption[Pure substance fugacity coefficients over temperature: --- 
Peng-Robinson EOS, -~-~- fit to the Peng-Robinson 
EOS, cf. equations (\ref{phin2}) to (\ref{phico2}).]{Vrabec et al.} 
\label{f2} 
\end{figure}

The fugacity coefficients, which describe the non-ideality of the vapor phase, were calculated with the Peng-Robinson EOS as well. 
The values do have a considerable temperature dependence, cf. Figure 
\ref{f2}, and were thus included in the present Henry model. 
Quadratic fits were found to be sufficient for all components, i.e. for N$_2$ 
\begin{eqnarray} 
\phi_{\rm N2}=5.5276-0.038326~T/{\rm K}+0.00008294~(T/{\rm K})^2,\label{phin2} 
\end{eqnarray}for O$_2$ 
\begin{eqnarray} 
\phi_{\rm O2}=0.54440-0.00053697~T/{\rm K}+0.00001082 
~(T/{\rm K})^2,\label{phio2} 
\end{eqnarray} 
and for CO$_2$ 
\begin{eqnarray} 
\phi_{\rm CO2}=0.51216+0.0046540~T/{\rm K}-0.00001082 
~(T/{\rm K})^2.\label{phico2} 
\end{eqnarray} 
Equations (12) to (19) define the present Henry model to describe the VLE of the ternary mixture in the CO$_2$-rich region.

\section{Results and Discussion} 
\label{rnd} 
In this section, present thermodynamic models are validated against experimental data. 
It is started with the three binary subsystems and subsequently the ternary case is regarded. 
Particular attention is given to the molecular model as it was used to predict data for development and validation of the Henry model.

\subsection{Nitrogen+Oxygen} 
 
\begin{figure}[ht] 
\epsfig{file=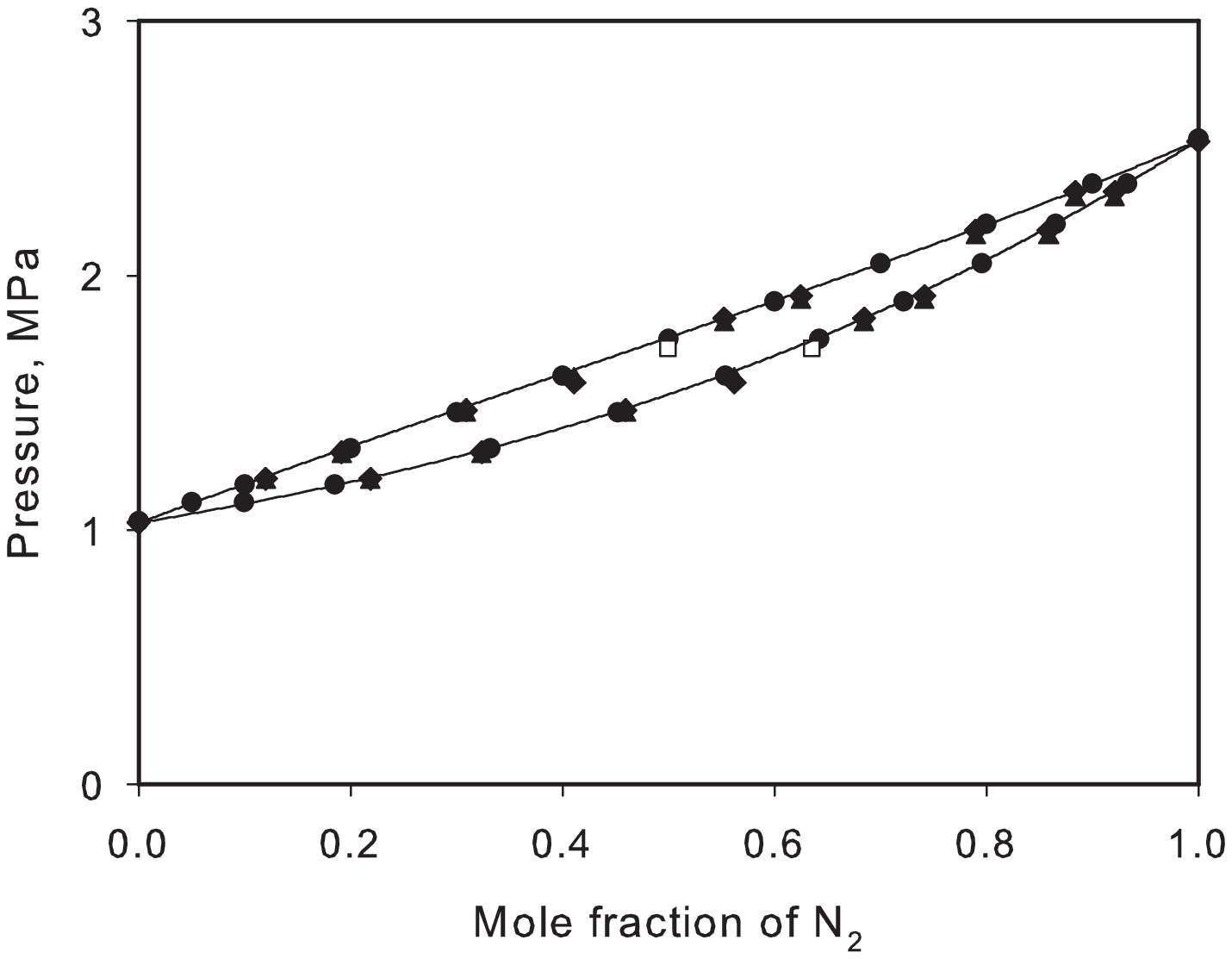,scale=0.95} 
\bigskip 
\bigskip 
\bigskip 
\bigskip 
\caption[Vapor-liquid equilibrium of the binary mixture N$_2$+O$_2$ at 
$-$153.15$~^{\circ}$C:  
$\bullet$, {\footnotesize $\blacklozenge$}, $\blacktriangle$ 
Experimental data \cite{dodg_no, kric_no , dodg1_no}, {$\square$} Simulation results, --- Peng-Robinson EOS.] {Vrabec et al.} 
\label{f3} 
\end{figure}

In Figure \ref{f3}, molecular simulation results and the Peng-Robinson EOS are compared to experimental VLE data \cite{dodg_no, kric_no , dodg1_no} of N$_2$+O$_2$ at $-$153.15$~^{\circ}$C. 
This temperature is considerably lower than the target range of $-$55 to $-$20$~^{\circ}$C due to the fact that both components have very low critical temperatures. 
It can be seen that the Peng-Robinson EOS correlates the experimental data well. By simulation, an equimolar composition in the liquid phase was regarded, where the mixing effect is strongest. 
The agreement between the three data sets is very good, which is also the case for other temperatures (not shown here). 
 
\subsection{Nitrogen+Carbon Dioxide} 
 
\begin{figure}[ht] 
\epsfig{file=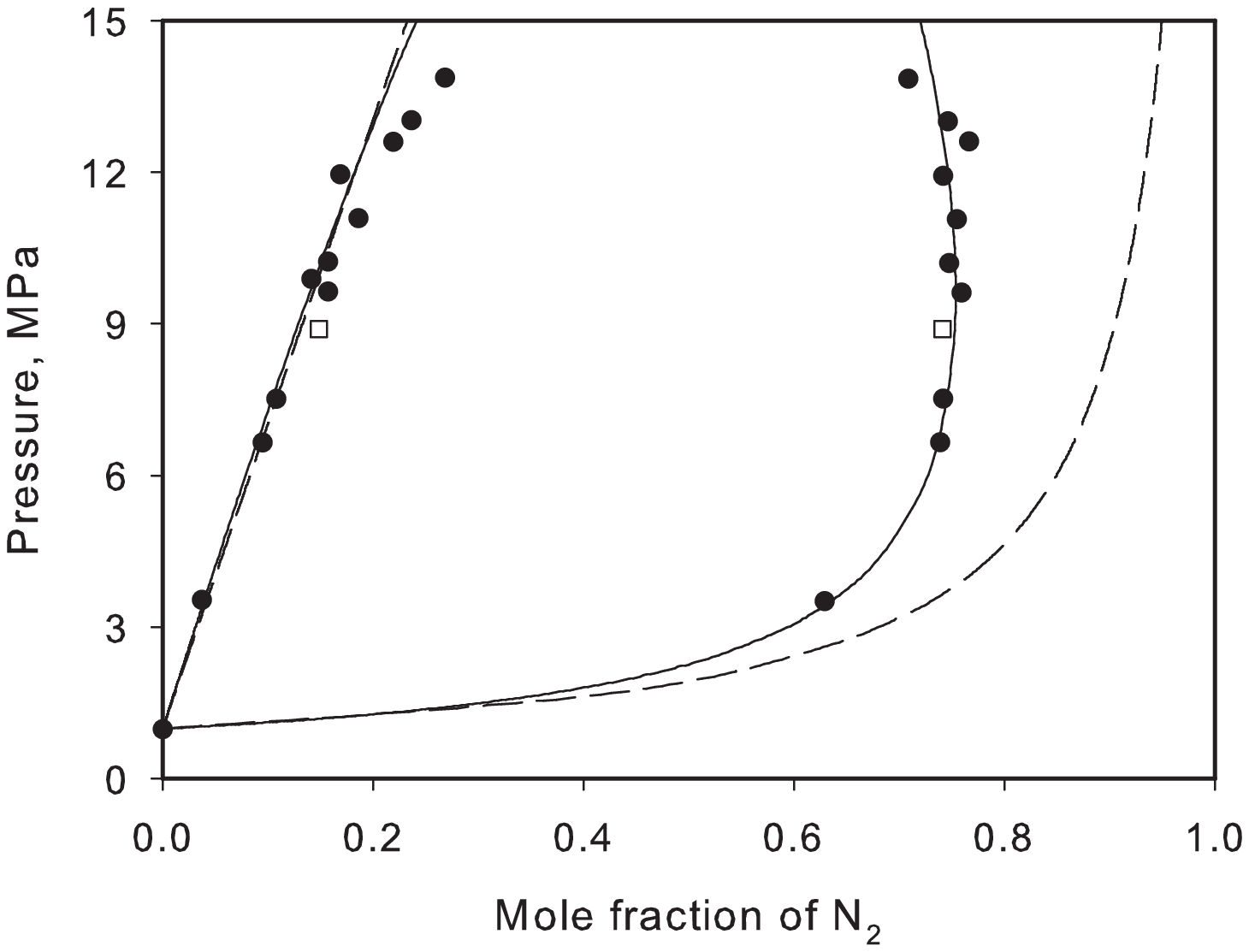,scale=0.95} 
\bigskip 
\bigskip 
\bigskip 
\bigskip 
\caption[Vapor-liquid equilibrium of the binary mixture N$_2$+CO$_2$ at $-$40.3$~^{\circ}$C:  
{\large $\bullet$} Experimental data \cite{zenn}, {$\square$} Simulation results,     
--- Peng-Robinson EOS, -~-~- Henry model.]{Vrabec et al.} 
\label{f4} 
\end{figure}

Figure \ref{f4} depicts the VLE of N$_2$+CO$_2$ at $- 40.3~^{\circ}$C including experimental data \cite{ zenn}, molecular simulation results, Peng-Robinson EOS, and Henry model. 
The experimental data shows some scatter, the simulation results are within this error bound. This also holds for the Peng-Robinson EOS for pressures up to 8 MPa. 
Approaching the critical region of the mixture, it is found that the Peng-Robinson EOS overshoots considerably which is a well known problem of this thermodynamic model. 
The Henry model is insufficient at 0$~^{\circ}$C (not shown here), but for lower temperatures, e.g. cf. Figure \ref{f4}, it agrees well in the CO$_2$-rich region. 
By closer inspection of the data, which will be made below, some deviations are found on the dew line that can hardly be seen with the resolution chosen for Figure \ref{f4}. 
 
\subsection{Oxygen+Carbon Dioxide} 
 
\begin{figure}[ht] 
\epsfig{file=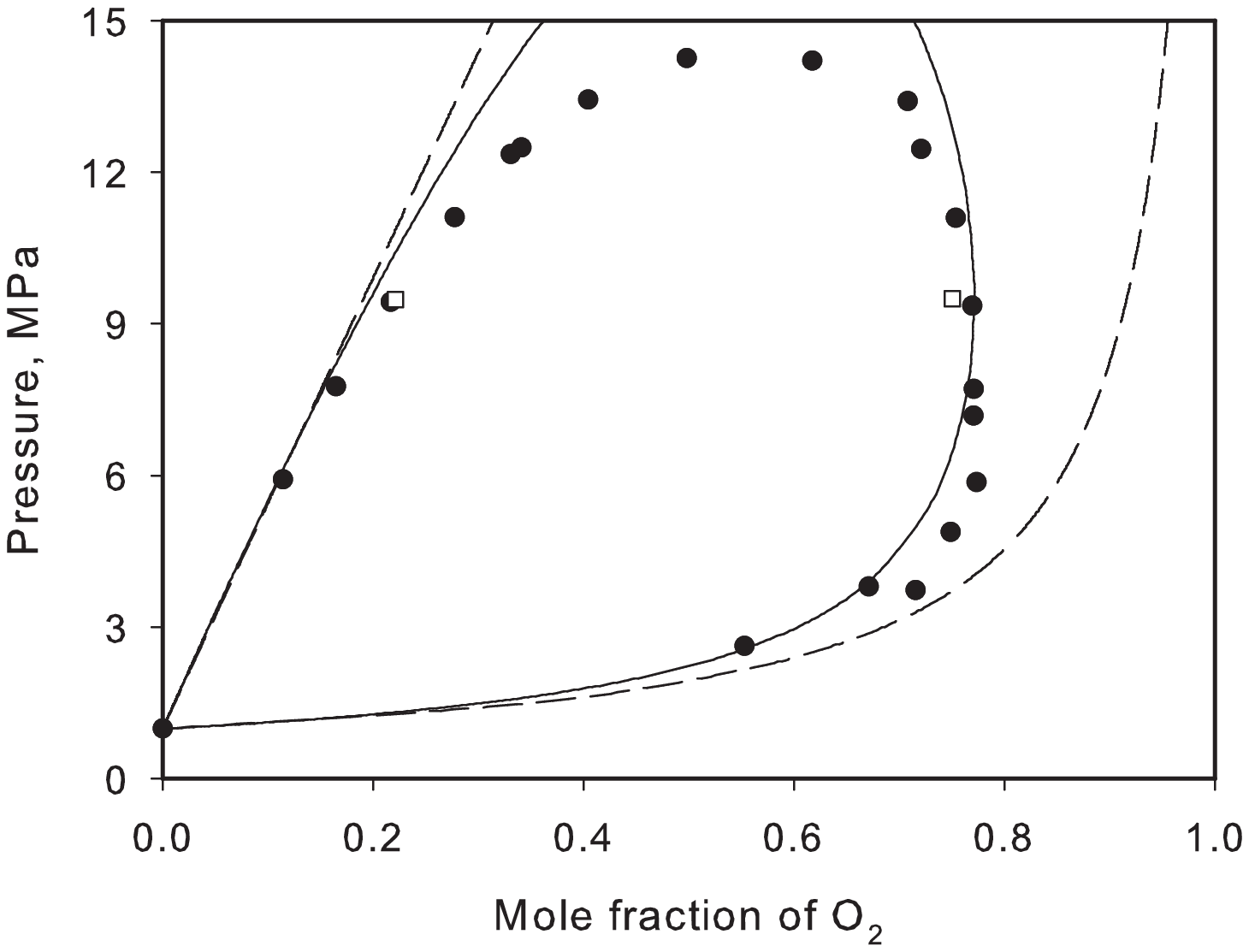,scale=0.95} 
\bigskip 
\bigskip 
\bigskip 
\bigskip 
\caption[Vapor-liquid equilibrium of the binary mixture O$_2$+CO$_2$ at $-$40.3$~^{\circ}$C:  
{\large $\bullet$} Experimental data \cite{fred}, {$\square$} Simulation results,     
--- Peng-Robinson EOS, -~-~- Henry model.]{Vrabec et al.} 
\label{f5} 
\end{figure}

Figure \ref{f5} presents the VLE data of O$_2$+CO$_2$ at $-$40.3$~^{\circ}$C from experiment  
\cite{fred}, molecular simulation, Peng-Robinson EOS, and Henry model. 
As before, the experimental data shows some scatter, particularly on the dew line. 
Again, it can be seen that the Peng-Robinson EOS overshoots in the critical region of the mixture. 
The agreement between Peng-Robinson EOS and Henry model is good in the CO$_2$-rich region. 
The molecular model shows reliable results, also for other temperatures (not shown here). 
As for the previous binary mixture, some deviations are observed on the dew line. 
 
\subsection{Nitrogen+Oxygen+Carbon Dioxide} 
 
\begin{figure}[ht] 
\epsfig{file=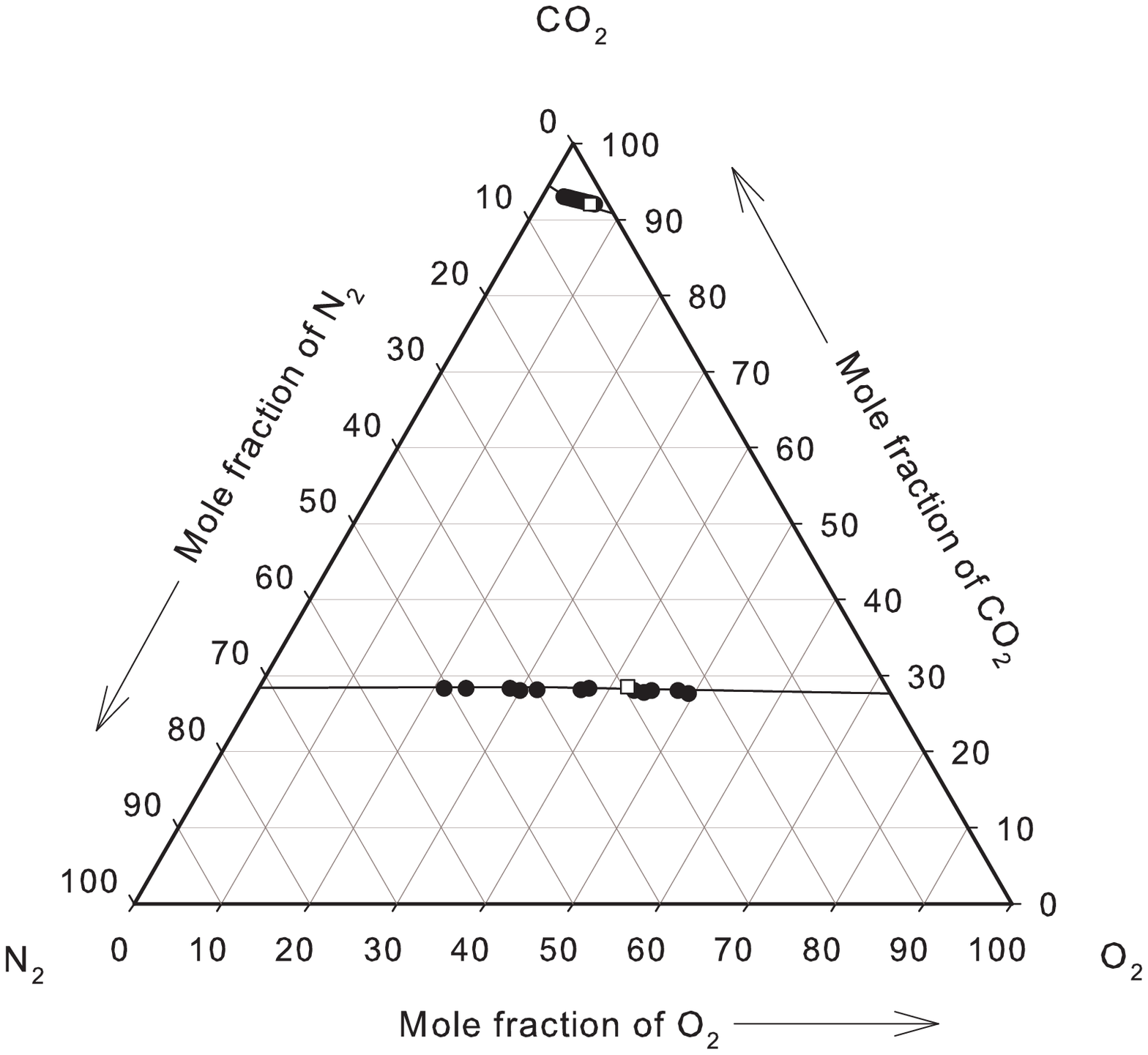,scale=0.95} 
\bigskip 
\bigskip 
\bigskip 
\bigskip 
\caption[Vapor-liquid equilibrium of the ternary mixture N$_2$+O$_2$+CO$_2$ at 
$-$40.3$~^{\circ}$C and 5.17 MPa:  
{\large $\bullet$} Experimental data \cite{zenn}, {$\square$} Simulation results,     
--- Peng-Robinson EOS.]{Vrabec et al.} 
\label{f6} 
\end{figure}

\begin{figure}[ht] 
\epsfig{file=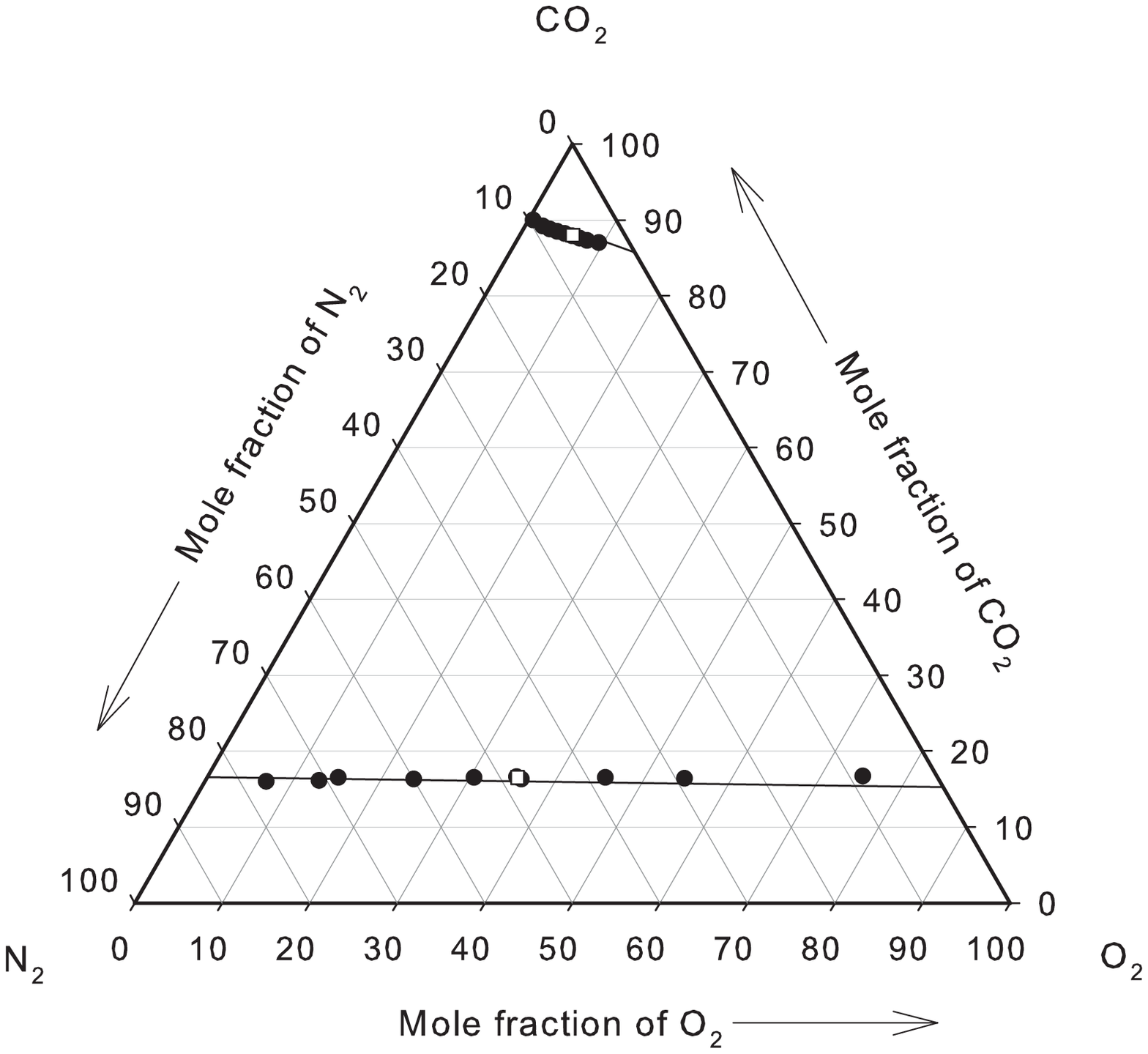,scale=0.95} 
\bigskip 
\bigskip 

\bigskip 
\bigskip 
\caption[Vapor-liquid equilibrium of the ternary mixture N$_2$+O$_2$+CO$_2$ at 
$-$55.0$~^{\circ}$C and 6.90 MPa:  
{\large $\bullet$} Experimental data \cite{zenn}, {$\square$} Simulation results,     
--- Peng-Robinson EOS.]{Vrabec et al.} 
\label{f7} 
\end{figure}

In Figures \ref{f6} and \ref{f7} simulation results and the Peng-Robinson EOS are compared with experimental ternary VLE data of the ternary system at $-$40.3 and $-$55$~^{\circ}$C \cite{zenn}. 
Due to the fact that experimental data is available only at high pressures between 5.17 and 13.10 MPa, the Henry model is not included into this validation. 
The Peng-Robinson EOS represents the VLE well at pressures of 6.9 MPa and below as depicted in Figures \ref{f6} and \ref{f7}. 
But it shows larger deviations at higher pressures approaching the critical region (not shown here). 
In the target region of state points, molecular simulation data is throughout in very good agreement with the experimental data. 
 
\subsection{Carbon Dioxide-Rich Region} 
\label{further} 
The present Henry model is developed for the design of technical applications in the CO$_2$-rich composition range. 
Therefore, further validations in this region, at state points where no experimental data is available, are presented here. 
The molecular model, being validated on the basis of experimental mixture VLE as discussed above, was used as the benchmark here. 
 
\begin{figure}[ht] 
\epsfig{file=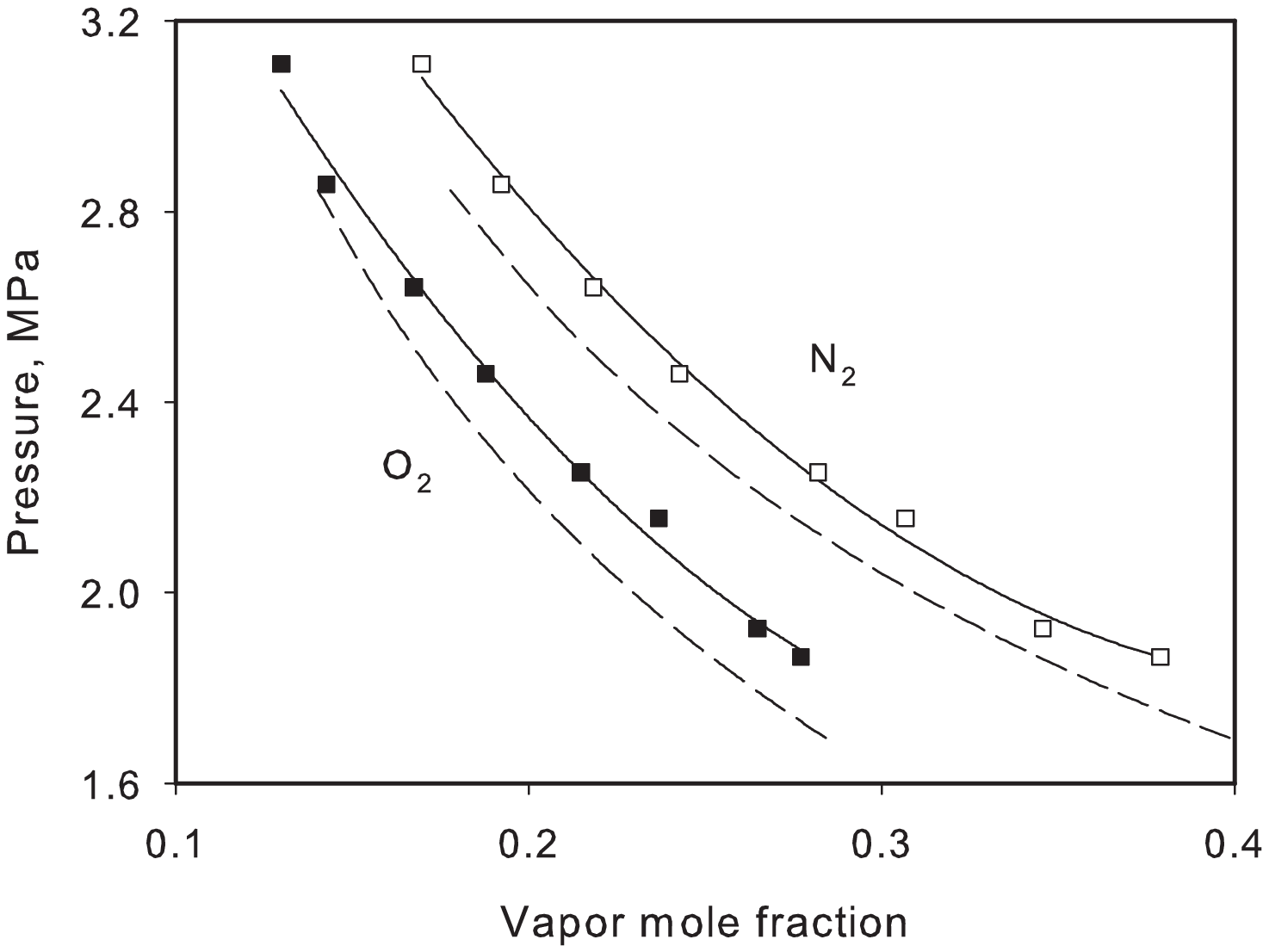,scale=0.95} 
\bigskip 
\bigskip 
\bigskip 
\bigskip 
\caption[Pressure over vapor mole fraction in vapor-liquid equilibrium at 
constant liquid mole fraction $x_{\rm N2}=x_{\rm O2}=0.01$ in the temperature range $-$55 to 
$-$20$~^{\circ}$C: {$\blacksquare$, $\square$} Simulation results,     
--- Peng-Robinson EOS, -~-~- Henry model.]{Vrabec et al.} 
\label{f8} 
\end{figure}

Figure \ref{f8} presents the pressure over the vapor mole fractions of N$_2$ and O$_2$ in VLE for constant liquid mole fractions $x_{\rm 
N2}$=$x_{\rm O2}$=0.01 in the temperature range from $-$55 to $-$20$~^{\circ}$C. 
Simulation data, Peng-Robinson EOS, and Henry model are compared. 
It can be seen that simulation data and Peng-Robinson EOS agree very well, deviations are minor. 
The Henry model deviates somewhat, yielding approximately 8\% too low pressures and 5\% too low vapor mole fractions. 

\begin{figure}[ht] 
\epsfig{file=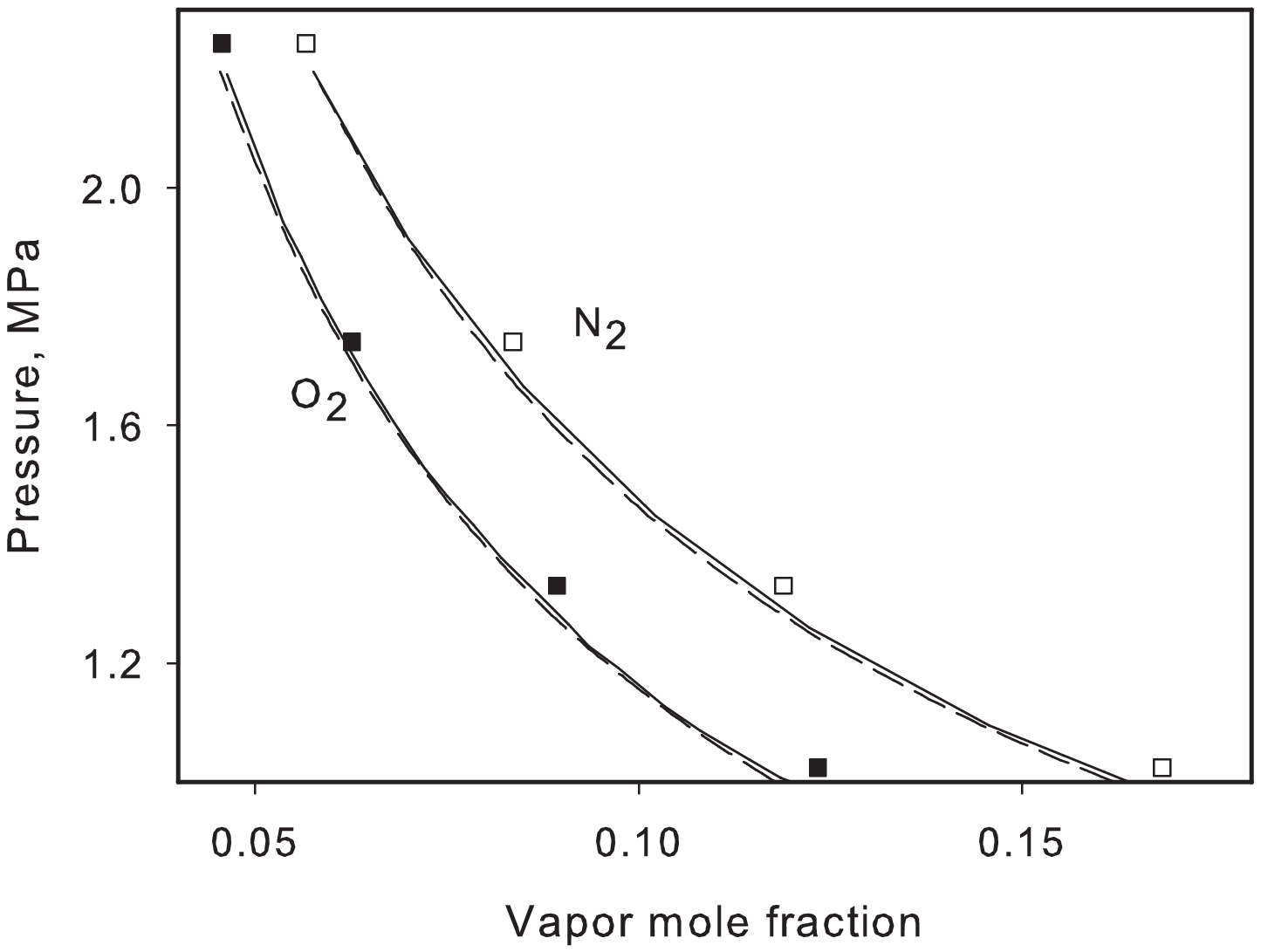,scale=0.95} 
\bigskip 
\bigskip 
\bigskip 
\bigskip 
\caption[Pressure over vapor mole fraction in vapor-liquid equilibrium at 
constant liquid mole fraction $x_{\rm N2}=x_{\rm O2}=0.0025$ in the temperature range $-$55 to 
$-$20$~^{\circ}$C: {$\blacksquare$, $\square$} Simulation results, --- Peng-Robinson EOS, -~-~- Henry model.]{Vrabec et al.} 
\label{f9}

\end{figure}

The limits, where the Henry model shows deviations of less than 2\% from molecular model and Peng-Robinson EOS, were investigated. 
Table \ref{max} shows that the Henry model is reliable for CO$_2$ liquid mole fractions above 0.995. 
This limit is examined in Figure \ref{f9}, where the pressure over the vapor mole fractions of N$_2$ and O$_2$ in VLE for constant liquid mole fractions $x_{\rm N2}$=$x_{\rm O2}$=0.0025 is shown in the temperature range from $-$55 to $-$20$~^{\circ}$C. 
The results from all three models agree well, proving the reliability of the thermodynamic models presented in this work.

\section{Conclusion} 
In this work three thermodynamic approaches, i.e.~a molecular model, the Peng-Robinson EOS, and a model based on Henry's law constants, were used to investigate the VLE of the ternary mixture N$_2$+O$_2$+CO$_2$ in the temperature range from $-$55 to $-$20$~^{\circ}$C. 
A thorough validation by comparison to experimental VLE data was made, where possible. 
The molecular model and the Peng-Robinson EOS are appropriate throughout, except in the critical region. 
For the very CO$_2$-rich region, which is important for purification processes, the computationally convenient Henry model can be used reliably.

\section{Acknowledgment} 
The authors thanks the Deutsche Bundesstiftung Umwelt for the financial support under the grant "Einsatz von CO$_2$ als K\"altemittel bei der 
CO$_2$-Verfl\"ussigung". The simulations were performed on the HP XC6000 super computer at the Steinbuch Centre for Computing, Karlsruhe under the grant MMSTP.

\clearpage
\noindent {\bf \large List of symbols} \\[-1cm]

\subsection*{Latin Letters}
\begin {tabbing}
\hspace{2.5cm}    \= \hspace{13cm}  \kill
   $ a       $ \> parameter of Peng-Robinson equation of state			\\[-0.12cm]
   $ A       $ \> coefficients of correlation 						\\[-0.12cm] 
   $ b       $ \> parameter of Peng-Robinson equation of state			\\[-0.12cm]
   $ f_Q      $ \> trigonometrical function depending on molecular orientations	\\[-0.12cm]
   $ H       $ \> Henry's law constant 						\\[-0.12cm]
   $ i       $ \> molecule index 							\\[-0.12cm]
   $ j       $ \> molecule index 							\\[-0.12cm]
   $ k_B     $ \> Boltzmann's constant, $k_B$ = 1.38066$\cdot10^{\rm 23}$ J/K 	\\[-0.12cm]
   $ k_{ij}  $ \> binary parameter of Peng-Robinson equation of state	\\[-0.12cm]
   $ L       $ \> elongation 							\\[-0.12cm]
   $m       $ \> molecular mass 							\\[-0.12cm]
   $ M      $ \> membrane mass parameter					\\[-0.12cm]
   $ n      $ \> exponent of correlation						\\[-0.12cm]
   $ p       $ \> pressure								\\[-0.12cm]
   $ Q       $ \> quadrupolar momentum						\\[-0.12cm]
   $ R       $ \> ideal gas constant							\\[-0.12cm]
  $ T       $ \> temperature							\\[-0.12cm]
   $ u       $ \> pair potential							\\[-0.12cm]
   $ v       $ \> molar volume							\\[-0.12cm]
   $ V      $ \> extensive volume							\\[-0.12cm]
   $ x       $ \> mole fraction in liquid phase					\\[-0.12cm]
   $ y       $ \> mole fraction in vapor phase					\\
\end{tabbing}

\subsection*{Greek Letters}
\begin {tabbing}
\hspace{2.5cm}    \= \hspace{13cm} \kill
   $ \gamma $ \> activity coefficient		 				\\[-0.12cm]
   $ \epsilon $ \> Lennard-Jones energy parameter 				\\[-0.12cm]
   $ \xi      $ \> binary interaction parameter 					\\[-0.12cm]
   $ \mu      $ \> chemical potential						\\[-0.12cm]
   $ \rho   $ \> number density		 					\\[-0.12cm]
   $ \sigma   $ \> Lennard-Jones size parameter 					\\[-0.12cm]
   $ \phi      $ \> fugacity coefficient	 					\\[-0.12cm]
   $ \psi      $ \> potential energy of test molecule				\\[-0.12cm]
   $ \omega   $ \> acentric factor						\\
\end{tabbing}

\subsection*{Subscripts}
\begin {tabbing}
\hspace {2.5cm}  \= \hspace{13cm} \kill
%
   $ a       $ \> count variable for molecule sites	\\[-0.12cm]
     A         \> related to component A		\\[-0.12cm]
   $ b       $ \> count variable for molecule sites	\\[-0.12cm]
     B         \> related to component B		\\[-0.12cm]
   $ c       $ \> critical value			\\[-0.12cm]
   $ i       $ \> related to component $i$		\\[-0.12cm]
   $ ij      $ \> related to components $i$ and $j$	\\[-0.12cm]
   $ j       $ \> related to component $j$		\\[-0.12cm]
   $ m       $ \> mixture				\\[-0.12cm]
     Q	       \> quadrupole				\\
\end{tabbing}

\clearpage
\subsection*{Superscripts}
\begin {tabbing}
\hspace {2.5cm}  \= \hspace{13cm} \kill
%
   $ \infty  $ \> at infinite dilution 	\\[-0.12cm]
   $ s     $ \> saturated 	\\
\end{tabbing}
\subsection*{Abbreviations}
\begin {tabbing}
\hspace {2.5cm}  \= \hspace{13cm} \kill
%
     2CLJ      \> two-center Lennard-Jones 				\\[-0.12cm]
     2CLJQ   \> two-center Lennard-Jones plus point quadrupole 		\\[-0.12cm]
     EOS      \> equation of state 						\\[-0.12cm] 
$NpT$       \> isobaric-isothermal ensemble 				\\[-0.12cm]
$NVT$       \> canonic ensemble                				\\[-0.12cm]
$\mu VT$ \> grand canonical ensemble 				\\[-0.12cm]
     VLE       \> vapor-liquid equilibria 					\\
\end{tabbing}
\subsection*{Vector properties}
\begin {tabbing}
\hspace{2.5cm}    \= \hspace{13cm} \kill
   $ {\bm r}_{ij}   $ \> center-center distance vector between two molecules $i$ and $j$\\[-0.12cm]
   $ {\bm \omega}   $ \> orientation vector of a molecule				\\ 
\end{tabbing} 
\clearpage


\clearpage

\begin{table}[ht] 
\noindent 
\caption[]{Parameters of the molecular models for the pure fluids, taken from  
\cite{vraquad}.} 
\label{2CLJDmodels} 
\bigskip 
\begin{center} 
\begin{tabular}{lcccc} \hline 
Fluid & $\sigma/$\r{A}  &  $\left(\epsilon/k_{\rm B}\right)/$K  &  $L/$\r{A}  & $Q/$D\r{A} \\ \hline
$\rm N_2$             			       & 3.3211 &34.897   & 1.0464      & 1.4397 \\  
$\rm O_2$             			       & 3.1062 &43.183   & 0.9699      & 0.8081 \\ 
$\rm CO_2$             			       & 2.9847 &133.22~~~   & 2.4176      & 3.7938 \\ \hline 
\end{tabular} 
\end{center} 
\end{table}

\begin{table}[ht] 
\noindent 
\caption[]{Binary interaction parameters of the molecular model, taken from \cite{vlemix}.} 
\label{xi} 
\bigskip 
\begin{center} 
\begin{tabular}{lc}  \hline 
Mixture 		& 	${\xi}$	\\ \hline
N$_2$ + O$_2$ 		& 	1.007	\\ 
N$_2$ +  CO$_2$		& 	1.041	\\ 
O$_2$ + CO$_2$		& 	0.979	\\ \hline 
\end{tabular} 
\end{center} 
\end{table} 
 
\begin{table}[ht] 
\noindent 
\caption[]{Pure substance parameters of the Peng-Robinson EOS, taken from 
\cite{refprop}.} 
\label{PRpure} 

\bigskip 
\begin{center} 
\begin{tabular}{lccccccc} \hline 
Fluid   &  $T_c$/K  &  $p_c$/MPa  &  $\omega$ \\ \hline
$\rm N_2$      & 126.19	&3.3958 & 0.0372 \\ 
$\rm O_2$      & 154.58	&5.0430 & 0.0222 \\ 
$\rm CO_2$     & 304.13	&7.3773 & 0.2239 \\ \hline 
\end{tabular} 
\end{center} 
\end{table} 
 
\clearpage 
 
\begin{table}[ht] 
\noindent 
\caption[]{Binary parameters of the Van der Waals one-fluid mixing rule 
adjusted in the present work.} 
\label{binpar} 
\bigskip 
\begin{center} 
\begin{tabular}{lr}  \hline 
Mixture 		& 	$k_{ij}$~~~~	\\ \hline
N$_2$ + O$_2$		& 	$-$0.0119   \\  
N$_2$ + CO$_2$ 		& 	0.0015	\\  
O$_2$ + CO$_2$		& 	0.124\0     \\ \hline 
\end{tabular} 
\end{center} 
\end{table} 
\begin{table}[ht] 
\noindent 
\caption[]{Parameters of the vapor pressure correlation for CO$_2$, taken from \cite{wagnerps}.}
\label{an} 
\bigskip 
\begin{center} 
\begin{tabular}{lrc}  \hline 
$i$ 		& 	$A_i$~~~~	&	$n_i$ \\ \hline
1 		& 	$-$6.95626	&	1	\\ 
2		& 	1.19695	&	3/2	\\
3		& 	$-$3.12614	&	3	\\ 
4		& 	2.99448	&	6	\\ \hline 
\end{tabular} 
\end{center} 
\end{table} 
 
\begin{table}[ht] 
\noindent 
\caption[]{Minimum mole fractions of CO$_2$ in the liquid and vapor where the 
Henry model deviates by less than 2 \% from the molecular model and the Peng-Robinson EOS.} 
\label{max} 
\bigskip 
\begin{center} 
\begin{tabular}{lcc}  \hline 

$T$/$^{\circ}$C	&$x_{\rm CO2}$	&$y_{\rm CO2}$\\ \hline
$-$20		& 0.994		&0.872	\\ 
$-$35 	& 0.995		&0.811	\\ 
$-$55		& 0.996		&0.706	\\ \hline 
\end{tabular} 
\end{center} 
\end{table} 
 
\clearpage

\listoffigures 
\clearpage 
 
 
\clearpage 
 
\clearpage 
 
\clearpage 
 
\clearpage 
 
\clearpage
  
\clearpage 
  
\clearpage 
 
\clearpage

\clearpage 
 
\end{document}